\documentstyle[12pt,epsf]{article}

\def\simge{\mathrel{%
   \rlap{\raise 0.511ex \hbox{$>$}}{\lower 0.511ex \hbox{$\sim$}}}}
\def\simle{\mathrel{
   \rlap{\raise 0.511ex \hbox{$<$}}{\lower 0.511ex \hbox{$\sim$}}}}

\newcommand{\beq}{\begin{equation}}

\newcommand{\eeq}{\end{equation}}

\newcommand{\al}{\alpha}

\newcommand{\de}{\delta}
\newcommand{\di}{\displaystyle}
\newcommand{\Ga}{\Gamma}
\newcommand{\ga}{\gamma}
\newcommand{\La}{\Lambda}
\newcommand{\la}{\lambda}

\newcommand{\si}{\sigma}

\newcommand{\ve}{\varepsilon}

\begin{document}

\begin{titlepage}
\setlength{\parskip}{0.25cm}
\setlength{\baselineskip}{0.25cm}
\begin{flushright}
DO-TH 99/24\\  
\vspace{0.2cm}
hep--ph/0002021
\\
\vspace{0.2cm}
February 2000
\end{flushright}
\vspace{1.0cm}
\begin{center}
{\LARGE{\bf{Leptogenesis with ``Fuzzy Mass Shell'' for Majorana Neutrinos}}}
\vspace{1.2cm}

{\large O.\ Lalakulich$^*$, E.A.\ Paschos$^{**}$ 
and M.\ Flanz$^{**}$}\\

\vspace{1.0cm}
\normalsize
{$^*$\it Faculty of Physics, Rostov State University}\\
{\it Rostov-on-Don, Russia}\\

\vspace{0.5cm}
\normalsize
{$^{**}$\it Institut f\"{u}r Physik, Universit\"{a}t Dortmund}\\ 
{\it D-44221 Dortmund, Germany} \\ \vspace{0.5cm}
{\it e--mail}: Paschos@hal1.physik.uni-dortmund.de\\  

\vspace{1.5cm}
\end{center}

\begin{abstract}
We study the mixing of elementary and composite particles.  In quantum
field theory the mixing of composite particles originates in the 
couplings of the constituent quarks and for neutrinos in self--energy
diagrams.  In the event that the incoming and outgoing neutrinos have
different masses, the self--energy diagrams vanish because energy is
not conserved but the finite decaying widths make the mixing possible.
We can consider the neutrinos to be ``fuzzy'' states on their mass shell
and the mixing is understood as the overlap of two wavefunctions.  
These considerations restrict the mass difference to be approximately
equal to or smaller than the largest of the two widths:
$|M_i-M_j|\simle max\{\Ga_i,\,\Ga_j\}$. 
\end{abstract}
\end{titlepage}

\section{Lagrangian with Majorana neutrinos.}

During the last few years a lot of attention was paid to the 
possibility of creating a baryon asymmetry  through leptogenesis.
The proposed schemes introduce heavy Majorana neutrinos with
CP--violating couplings,  where both the so--called ``direct'' and
``indirect'' contributions to leptogenesis were considered.
In calculating the lepton asymmetry, the authors consider the
Standard Model (SM) with the usual particle content plus three
Majorana neutrinos, which are singlets under the weak SU(2)--group
\cite{fuku} - \cite{roulet}.

The part of the Lagrangian with Majorana neutrinos consists of the 
Majorana mass term and the Yukawa interactions of these neutrinos with
leptons and Higgs bosons:
\beq
L=\frac12\sum_i M_i\overline{N_i}\, N_i+ 
 \sum_{\al,\,i}h_{\al i}\overline{l_{L\al}}\,\phi\, P_R N_i
      +\sum_{\al,\,i}h^*_{\al i} \overline{N_i} P_L\, l_{L\al} \phi^+\,
      + h.c.
\label{inilag}
\eeq
In this Lagrangian $l_{L\al}$ are left-handed lepton doublets of the SM,
$\phi$ is the Higgs doublet of the SM, $\al,i=1,2,3$ 
denote the index of fermion generation and
$N_i$ is the self-conjugate Majorana field:
\begin{equation}
N_i=N_{Ri}+(N_{Ri})^c=\left(\begin{array}{c} 0 \\ N_{Ri} \end{array} \right)+
            \left(\begin{array}{c} (N_{Ri})^c \\ 0 \end{array} \right)
=\left(\begin{array}{c} (N_{Ri})^c \\ N_{Ri} \end{array} \right).   
\end{equation}
One should remember here that $(N_{Ri})^c$ is a left-handed antiparticle
$(N_{Ri})^c=(N_i^c)_L$.

Another (more explicit) way of writing this Lagrangian is 
used in \cite{weiss}:
\beq
\begin{array}{ll} \di
L=\sum_{i} M_i\left[\overline{(N_{Ri})^c} N_{Ri}+ 
            \overline{N_{Ri}} (N_{Ri})^c \right]
+\sum_{\al,\,i}h_{\al i}\overline{l_{L\al}} N_{Ri}\phi 
+\sum_{\al,\,i}h^*_{\al i} \overline{N_{Ri}} l_{L\al} \phi^+
\\[4mm] \di \hspace{40mm}
      +\sum_{\al,\,i}h_{\al i} \overline{(N_{Ri})^c} (l_{L\al})^c \phi
+\sum_{\al,\,i}h^*_{\al i}\overline{(l_{L\al})^c} (N_{Ri})^c \phi^+\,.
\end{array}
\label{pasclag}
\eeq
One can easily show that the two Lagrangians (\ref{inilag}) 
and (\ref{pasclag}) are identical. In fact, 
by the definition of charge--conjugate fields one can show that  
\begin{equation}
\overline{(l_{L\al})^c} \phi^+(N_{Ri})^c
=-(l_{L\al})^TC^{-1}\phi^+C \overline{N_{Ri}}{}^T 
=-(l_{L\al})^T \phi^+ \overline{N_{Ri}} {}^T 
=\overline{N_{Ri}}\phi^+l_{L\al}. 
\end{equation}

Notice that the mass term in (\ref{inilag}) violates lepton number 
by two units and the Yukawa interaction terms violate the CP-symmetry.
In general, it  is desirable to have a gauge theory
with a symmetry responsible for lepton number conservation. In such
a theory, the
Majorana mass term is generated as the result of spontaneous
breaking of lepton number.  This theory can be
the goal of future investigations; meanwhile, in the model under 
consideration, the asymmetry  
is believed to be generated at temperatures bigger than the electroweak 
symmetry 
scale, but lower than the scale where the Majorana mass is
created. 
Thus the Lagrangian (\ref{inilag}) is an
``intermediate--energy'' effective Lagrangian. At the high energies 
considered here,
the vacuum expectation value for the Higgs  condensate is very small,
so that the masses (here we mean vacuum  masses and neglect temperature
contributions) of the charged leptons $m_{\al}$ and Higgs particles 
$m_{\phi}$ are negligibly small or zero.

%
\section{Indirect contribution to leptogenesis.} 

The Lagrangian  (\ref{inilag}) was considered several times 
\cite{fuku} - \cite{buch},
where the so-called ``direct''  contribution to leptogenesis
was computed.  Here we are interested in 
the ``indirect'' contribution to leptogenesis, calculated in references
\cite{eap}--\cite{roulet} and reviewed recently in \cite{ranga}.


\begin{figure}[h]
\begin{center}
\hspace{2cm}
\epsffile{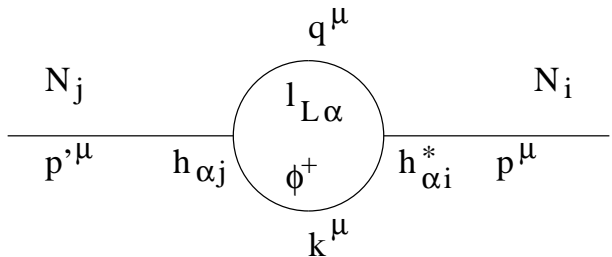}
\end{center}
\caption{The self-energy diagram of the heavy Majorana neutrino}
\end{figure}

The indirect contribution is described by the self-energy diagram in Fig.1.
This diagram was investigated in the context of two different approaches.
In the first one
the self--energy diagram is considered to be the {\it intermediate} 
state of
some physical process \cite{apost,apint}
(for example, lepton--Higgs scattering \cite{flanz}).
In this case the four-momentum of the Majorana neutrino is off the 
mass shell
and is  determined by the four-momenta  of the incoming (outgoing)  
lepton and
Higgs boson. We wish to emphasize is that  there are no  
restrictions on the value of neutrino four-momentum. 
In this approach the lepton asymmetry 
appears in the scattering process $l^c\phi\to l\phi^+$, through 
interference of the tree-level diagram with the one-loop 
diagram \cite{ranga}.

The second approach, presented earlier in \cite{eap,weiss}, uses the
notions of transition $N_j\to N_i$, 
originating from 
the one--loop diagram mentioned already.
If $i=j$ then the diagram is analogous to any self-energy diagram and
describes the correction to the fermion (Majorana neutrino in our case)
mass. The precise role and calculation of this sort of diagrams 
is described in textbooks (see, for example, corrections to the electron 
mass in \cite{zuber,wein}.

A new situation appears for $i\neq j$.  Now we have a free stable
particle with mass $M_i$ as incoming particle and it is impossible
to transform it into a stable particle with mass $M_j$, because the 
process does not conserve energy and momentum.  This transition can
happen for unstable particles provided their mass difference is
comparable to their widths.  In addition the Majorana states are in
a thermal bath interacting with the other fields so that energy and
momentum is continuously exchanged with the background fields.  
If the transition probabilities are smooth functions of energies and
widths, then
it is a good approximation to calculate the transition amplitudes
and probabilities per unit time and volume for unstable Majorana
neutrinos and then introduce the results into the Boltzmann equations.
In the latter step we also introduce the particle densities for initial
and final states, as dictated by the thermodynamics of the early
universe.  A prototype of the thermodynamic calculation is described
in ref.\ \cite{flanz}, \cite{micha} and recent articles \cite{miele,fred}.  
The interaction with the background fields is an
integral part in the generation of the lepton asymmetry.  In this
article we calculate how the transition probabilities and the asymmetry
are generated for unstable particles. 

Before we address this topic, we review how the mixing occurs in a 
few known cases.  The mixing of the $K^0$ with $\bar{K}\, ^0$ is
described by the box--diagram.  In this case the $K^0$--mesons are
treated as composite particles made up of quarks.
The mixing diagrams are computed at the quark level, which are the 
fields of the basic Lagrangian (Standard Model).  In the diagrams
appear vertices of the form $\bar{u}\, s\, W$, because
the quarks mix in the Lagrangian through the 
Cabibbo--Kobayashi--Maskawa matrix.  
Thus the mixings in the above case originate 
from the mixings at vertices between quarks of different charges
and manifest themselves as the mixing of mesonic states through the 
interaction of their
constituent quarks.  In the above case energy and momentum is conserved.
For the mesonic states the mass difference is comparable to the 
widths.  For K--mesons 
\begin{displaymath}
 |\Delta M_K| = \frac{1}{2}|\Delta \Gamma_K|\sim \frac{1}{2}\Gamma_K
\end{displaymath}
and for the B--mesons \cite{pdg}
\begin{displaymath}
	\left( \frac{\Delta M}{\Gamma}\right)_{B_d} = 0.734\pm 0.035\, .
\end{displaymath}

The neutrinos have charged current couplings to the leptons where
the lepton number is conserved to a large degree of accuracy.  The
Majorana couplings, on the other hand, may have lepton number violating
charged couplings as seen in the Lagrangian of eqs.\ (1) or (3).  The
mass matrix of Majorana neutrinos in eq.\ (1) can be made real and 
at the classical (tree) level the generations do not mix.  
However, it
was noted in ref.\ \cite{weiss} that the one--loop effects can mix the 
Majorana states, i.e. the states defined at tree level are not the
physical ones.  To state the result in another way, we find that
Majorana neutrinos of different generation mix, while the electron
and the muon do not mix.  

In this paper we show that the instability of the Majorana neutrinos
is of principal importance for the mixing phenomenon.
The neutrinos have finite widths and can be treated, through the
uncertainty principle, as ``fuzzy states'' on their mass shell.  This
allows us to define the mixing of physical states in the case when
the difference of the tree--level neutrino masses is less or about
equal to their widths:
\beq
|M_i-M_j|\simle max\{ \Gamma_i,\; \Gamma_j \}.
\label{1} 
\eeq
Condition (5) can evidently be applied to any other particles.
For electron and muon, however,  $m_{\mu}-m_e\gg \Gamma_{\mu}$ and
consequently the mixing is very small.

\section{S-matrix for the mixing of unstable particles.}  

The general S-matrix theory necessarily uses the notion of
asymptotic states. This means that states of
incoming, $|in(t\to -\infty)\rangle$, and outgoing, 
$|out(t\to +\infty)\rangle$, particles are defined as states 
of free (noninteracting) particles at infinite times.
The second principal 
element of the $S-$matrix theory presupposes that interactions do 
not exist at all ``infinite'' times, but are ``turned on adiabatically'' 
from $t\to -\infty$ to $t=0$ and also ``turn off'' adiabatically
from $t=0$ to $t\to +\infty$. 
One, of course, understands that ``infinite'' here means macroscopically 
large in comparison to the time of interaction.  
This general argument can naturally be applied to  the case of 
unstable  particles.  In this case the time of ``turning 
on'' and ``off'' the interaction is easily estimated to be of the order of the 
inverse particle width.  We try to express this in a phenomenological
definition (the idea is similar to that of Breit and
Wigner when considering the cross section at resonance) of field 
operators for unstable particles.  For neutrinos we
introduce 
\beq \di
N_i(x)=\sum\limits_{\vec{k}, \la}\left[ 
u^i(\vec{k}, \la)a^i_{\vec{k}, \la}e^{-i\sqrt{\vec k\,{}^2+M_i^{2}}t}e^{i\vec k
\vec x} + v^i(\vec{k}, \la)a^{i+}_{\vec{k}, \la}e^{i\sqrt{\vec
k\,{}^2+M_i^{2}}t}  e^{-i\vec k \vec x}
\right] e^{-\Gamma_i|t|} 
\label{neutrino}
\eeq
The indices $i=1,2$ denote the generation of the neutrino and for the sake 
of simplicity  we consider only two generations.
$a_{\vec{k}, \lambda}^{i+}$ and $a_{\vec{k}, \lambda}^i$ are the fermionic
creation and annihilation operators.
Formula  (\ref{neutrino}) reflects the fact that 
neutrinos ``disappear'' (decay)
for infinite time ($N_i\to 0$ at $t\to\pm\infty$).
We assume that the widths are generated by the Lagrangian in eq.\ (1) and
calculated by the decay diagrams or from the absorptive part of the
diagonal terms of the self--energy.  Eq.\ (6) is the only step which
is not proved from field theory, since field theory does not consider
asymptotic states.  We think, however, that the physical meaning of
eq.\ (6) is rather clear and will help us understand the mixing of
Majorana neutrinos.  

In this paper we will follow the idea of an earlier paper \cite{weiss} 
and will introduce an ``effective Hamiltonian'' with non-diagonal 
terms in the mass matrix.  Initially they are calculated as elements of the
S-matrix to second order in the Yukawa--couplings by using perturbation
theory.
Let $i$ and $j$ be neutrinos of definite flavors, then
\beq
S_{ij}=-\frac12\int d^4x\int d^4x'\langle i|L_{int}(x)L_{int}(x')|j\rangle,
\label{sij}
\eeq
with the neutrino fields defined by eq.\ (\ref{neutrino}). 

\section{Loop calculation with unstable Majorana neutrinos.}
As mentioned in the previous section, we shall use perturbation theory.
To second 
order in the Yukawa couplings the $S-$matrix  element, corresponding
to the one-loop diagram, is given by (\ref{sij}).
>From the product of the interaction Lagrangians 
\[ \begin{array}{l} \di
\langle i |
\left\{ h_{\al\,m}\overline{l_{L\al}}(x)\phi(x)P_R N_m(x)+
       h_{\al\,m}^* \overline{N_m}(x)P_L \phi^+(x)l_{L\al}(x) \right\}
\times
\\[3mm] \hspace*{40mm} \di
\times
\left\{ h_{\al\,n}\overline{l_{L\al}}(x')\phi(x')P_R N_n(x')+
       h_{\al\,n}^* \overline{N_n}(x')P_L \phi^+(x')l_{L\al}(x') \right\}
|j\rangle
\end{array}
\]
two terms contribute to (\ref{sij}):
\beq 
\begin{array}{l} \di
2 h_{\al\,i}^* h_{\al\,j}\cdot\overline{N_i}(x) P_L \phi^+(x) l_{L\al}(x)
  \cdot\overline{l_{L\al}}(x')\phi(x')P_R N_j(x')
\\[3mm] \hspace*{40mm} \di
+2 h_{\al\,i} h_{\al\,j}^*\cdot P_R N_i(x) \phi(x) \overline{l_{L\al}}(x)
    \cdot\overline{N_j}(x') P_L \phi^+(x') l_{L\al}(x').
\end{array}
\label{two-terms}
\eeq
The new aspect of our calculation is the form of the field operators
for unstable neutrinos given in eq.\ (6).  We calculate in detail the
first term, using Wick's expansion
for the product of operators
\beq 
\begin{array}{l} 
\di
S_{ij}=\frac12\int d^4x\int d^4x' 2 h_{\al\,i}^*h_{\al\,j} \bar u^i_{p\si}
P_L e^{i\sqrt{\vec p\,{}^2+M_i^2}t-\Gamma_i|t|-i\vec p\vec x } 
\int\frac{d^4q}{(2\pi)^4} \frac{e^{-iq(x-x')}}{\not q-m_{\al}+i\ve}
\\[5mm] \di \hspace*{10mm}
\int\frac{d^4k}{(2\pi)^4} e^{-ik(x-x')}\left(\frac{1}{k^2-m_{\phi}^2+i\ve}-
\frac{1}{k^2-\Lambda^2+i\ve}\right) 
e^{-i\sqrt{\vec p {}\, '{}^2+M_j^2}t'-\Gamma_j|t'|+i\vec p{}\, ' \vec x '}
\!P_R u^j_{p'\si'},
\end{array} 
\eeq 
where $\Lambda$ is a parameter for the Pauli--Villars 
regularization.\footnote{Another regularization is equally possible.}
After integration over $d^3x$, $d^3k$, $d^3x'$ one obtains:
\beq 
\begin{array}{l} 
\di
S_{ij}=\frac12\int dt\int dt' 2 h_{\al\,i}^*h_{\al\,j} \bar u^i_{p\si}P_L
\int\frac{d^4q}{(2\pi)^4} \frac{\not q+m_{\al}}{q^2-m^2_{\al}+i\ve}
\int\frac{d k^0}{2\pi}\frac1{k^0{}^2-(\vec p-\vec q)^2-m_{\phi}^2+i\ve}
\\[5mm] \di \hspace*{10mm}
e^{i(\sqrt{\vec p\,{}^2+M_i^2}-k^0-q^0)t-\Gamma_i|t|}
 e^{i(\sqrt{\vec p\, {}' {}^2+M_j^2}-k^0-q^0)t'-\Gamma_j|t'|}
\delta(\vec p-\vec p \, ')(2\pi)^3 P_R u^j_{p'\si'}.
\end{array} 
\label{kolok}
\eeq
with a similar expression for the terms with the Pauli-Villars regularization
parameter.
An important feature of this formula is the product of the two integrals
\beq
J_1=\int dt \; e^{i(\sqrt{\vec p\,{}^2+M_i^2}-k^0-q^0)t-\Gamma_i|t|}
\qquad
J_2=\int dt' \, e^{i(\sqrt{\vec p\, '{}^2+M_j^2}-k^0-q^0)t'-\Gamma_j|t'|}.
\eeq
\noindent In the case of vanishing widths $\Gamma_i=\Gamma_j=0$ each 
of the integrals
is equal to a $\delta-$function and their product 
$\de(\sqrt{\vec p\,{}^2+M_i^2}-k^0-q^0)
                \de(\sqrt{\vec p\,{}^2+M_j^2}-k^0-q^0)$
is zero except when $M_i=M_j$ (with $\vec{p} = \vec{p} \, '$). 
This result is a mathematical demonstration 
of physical arguments, discussed in section 2. 
On the other hand, for non--zero widths the integrals $J_1$ and $J_2$ lead 
to  ``bell-shaped'' Lorentzian functions:
\beq
J_1\cdot J_2=(2\pi)^2 
    \frac{\Ga_i/\pi}{\Ga_i^2+(\sqrt{\vec p\,{}^2+M_i^2}-k^0-q^0)^2}\cdot
    \frac{\Ga_j/\pi}{\Ga_j^2+(\sqrt{\vec p\,{}^2+M_j^2}-k^0-q^0)^2}.
\label{J12}
\eeq
This product is non--zero and large when condition (\ref{1}) is satisfied;
its non--zero value can be graphically understood as the ``overlap''
of two ``bells''. So the limit $|M_i-M_j|\gg \Gamma_{i(j)}$, which is 
often used in the calculation of asymmetry, leads to very small overlap 
functions.
Another remark concerns the three-dimensional 
$\delta-$function with zero argument 
$(2\pi)^3\de(\vec p-\vec p \, ')=(2\pi)^3\de(\vec 0)$, which appears in 
eq.\ (10). We will keep it throughout 
the section and it corresponds to the volume element where the 
interaction takes place.

When we substitute (\ref{J12}) into (\ref{kolok}), and in addition
integrate
over  $k^0$, we obtain the first order expression for $\Gamma_{i(j)}$:
\beq
\begin{array}{l} \di
S_{ij}=h_{\al\,i}^*h_{\al\,j} 2\pi \delta(\vec p-\vec p)(2\pi)^3
\bar u^i_{p\si}P_L
\int\frac{d^4q}{(2\pi)^4} \frac{\not q+m_{\al}}{q^2-m^2_{\al}+i\ve} 
P_R u^j_{p\si'}
\\[5mm] \hspace*{30mm}  \di
\Biggl\{\frac{\di\Ga_j}{\di\pi} \frac{1}{(\sqrt{\vec p\,{}^2+M_i^2}
                      -\sqrt{\vec p\,{}^2+M_j^2})^2+\Ga_j^2}\times
\\[5mm] \hspace*{10mm}  \di
\times\left(\frac{1}{(\sqrt{\vec p\,{}^2+M_i^2}-q^0)^2
             -(\vec p-\vec q)^2-m_{\phi}^2}
-\frac{1}{(\sqrt{\vec p\,{}^2+M_i^2}-q^0)^2
             -(\vec p-\vec q)^2-\La^2} \right)
\\[10mm] \hspace*{30mm}  \di
+\frac{\di\Ga_i}{\di\pi} \frac{1}{(\sqrt{\vec p\,{}^2+M_j^2}
                      -\sqrt{\vec p\,{}^2+M_i^2})^2+\Ga_i^2}\times
\\[5mm] \hspace*{10mm}  \di
\times\left(\frac{1}{(\sqrt{\vec p\,{}^2+M_j^2}-q^0)^2
             -(\vec p-\vec q)^2-m_{\phi}^2}
-\frac{1}{(\sqrt{\vec p\,{}^2+M_j^2}-q^0)^2
             -(\vec p-\vec q)^2-\La^2} \right) \Biggr\}
\end{array}
\eeq
This expression, as is easily seen, reproduces the usual expression
for the loop integral after one introduces the 
zero-component of four momentum $p^0$:
\beq
\begin{array}{l} \di
S_{ij}=h_{\al\,i}^*h_{\al\,j} 2\pi \delta(\vec p-\vec p)(2\pi)^3
\bar u^i_{p\si}P_L
\int\frac{d^4q}{(2\pi)^4} \frac{\not q+m_{\al}}{q^2-m^2_{\al}+i\ve}
P_R u^j_{p\si'}\times
\\[5mm] \hspace*{10mm}  \di
\times\left\{
\frac{\Ga_j}{\pi}\frac{1}{(p^0-\sqrt{\vec p\,{}^2+M_j^2})^2+\Ga_j^2}
\frac{1}{(p^0-q^0)^2-(\vec p-\vec q)^2-m_{\phi}^2}
     \left|{}^{p^0=\sqrt{\vec p\,{}^2+M_i^2}}\right.\right.
\\[5mm] \hspace*{15mm}  \di
\left.
+\frac{\Ga_i}{\pi}\frac{1}{(p^0-\sqrt{\vec p\,{}^2+M_i^2})^2+\Ga_i^2}
\frac{1}{(p^0-q^0)^2-(\vec p-\vec q)^2-m_{\phi}^2}
     \left|{}^{p^0=\sqrt{\vec p\,{}^2+M_j^2}}\right.\right\}
\end{array}
\eeq
Integrals of this type are standard and the result for the limit
$m_{\al}\to 0$, $m_{\phi}\to 0$ is well known:
\[  
\begin{array}{c}
\di
Int=\lim\limits_{m_{\al},m_{\phi}\to 0}
\int\frac{d^4q}{(2\pi)^4} \frac{\not q+m_{\al}}{q^2-m^2_{\al}+i\ve}
\left(\frac{1}{(p^{\mu}-q^{\mu})^2-m_{\phi}^2+i\ve}-
      \frac{1}{(p^{\mu}-q^{\mu})^2-\La^2+i\ve}\right)= 
\\[7mm] \di
=-i\ga^{\mu}p_{\mu}\left(g_{dis}-\frac{i}{2}g_{abs}\right), 
\qquad \qquad
 g_{dis}=-\frac1{16\pi^2}\left(\frac12\ln\frac{\La^2}{p^2}+\frac34\right),
\qquad g_{abs}=\frac1{16\pi} 
\end{array}
\]
  It is recognized that the dispersive part  of this integral ($g_{dis}$)
can be reabsorbed in the definition of coupling constants,
while only the absorptive part ($g_{abs}$) survives and has physical
consequences; for more details see \cite{zuber,wein} and references therein.

Now the matrix element $S_{ij}$ is given by
\beq
\begin{array}{l} \di
S_{ij}=h_{\al\,i}^*h_{\al\,j} 2\pi \delta(\vec p-\vec p)(2\pi)^3
(-i)  \times
\\[5mm] \di \hspace{3mm} 
\times \left\{ \bar u^i_{p\si}P_L 
\left[ \ga^0(\sqrt{\vec p\,{}^2+M_i^2})-\vec\ga\vec p\,\right]
    \frac{\Ga_j}{\pi}\frac{-i/2\cdot g_{abs}}{(\sqrt{\vec p\,{}^2
       +M_i^2}-\sqrt{\vec p\,{}^2+M_j^2})^2+\Ga_j^2} P_R u_{p\si'}^j
\right. 
\\[5mm] \di \hspace{6mm} 
+ \left. \bar u^i_{p\si}P_L 
\left[ \ga^0(\sqrt{\vec p\,{}^2+M_j^2})-\vec\ga\vec p\,\right]
    \frac{\Ga_i}{\pi}\frac{-i/2\cdot g_{abs}}{(\sqrt{\vec p\,{}^2
       +M_j^2}-\sqrt{\vec p\,{}^2+M_i^2})^2+\Ga_i^2}P_R u_{p\si'}^j
\right\}
\end{array}
\eeq
We transform this expression with the help of the Dirac equation 
\beq
\begin{array}{l} \di
\bar u^i_{p\si}P_L 
 \left[ \ga^0\sqrt{\vec p\,{}^2+M_i^2}-\vec\ga\vec p\,\right]
=\bar u^i_{p\si}P_R M_i, 
\\[5mm] \di
\left[ \ga^0\sqrt{\vec p\,{}^2+M_j^2}-\vec\ga\vec p\,\right]
P_R\, u^j_{p\si'}=
 M_j P_L\, u^j_{p\si'},
\\[5mm] \di
\bar v^j_{p\si'}P_L 
 \left[ \ga^0\sqrt{\vec p\,{}^2+M_j^2}-\vec\ga\vec p\,\right]
=-\bar v^j_{p\si'}P_R M_j, 
\\[5mm] \di
\left[ \ga^0\sqrt{\vec p\,{}^2+M_i^2}-\vec\ga\vec p\,\right]
P_R\, v^i_{p\si}=-M_i P_L\, v^i_{p\si}.
\end{array}
\eeq
Finally we arrive at
\beq
\begin{array}{l} \di
S_{ij}^{(I)}=h_{\al\,i}^*h_{\al\,j} 2\pi \delta(\vec p-\vec p)(2\pi)^3
(-i) \frac12 \bar u^i_{p\si} \left\{ 
M_i \frac{\Ga_j}{\pi}\frac{-i/2\cdot g_{abs}}{(\sqrt{\vec p\,{}^2
       +M_i^2}-\sqrt{\vec p\,{}^2+M_j^2})^2+\Ga_j^2} 
\right. 
\\[6mm] \di \hspace{30mm} 
+ \left. M_j
    \frac{\Ga_i}{\pi}\frac{-i/2\cdot g_{abs}}{(\sqrt{\vec p\,{}^2
       +M_j^2}-\sqrt{\vec p\,{}^2+M_i^2})^2+\Ga_i^2}
\right\} u_{p\si'}^j
\end{array}
\eeq
This is the final expression for the S--matrix element originating
from the first term in eq. (\ref{two-terms}). The second term from eq.
(\ref{two-terms})  leads us to the similar
expression with $v-$spinors 
\beq
\begin{array}{l} \di
S_{ij}^{(II)}=h_{\al\,i}h_{\al\,j}^* 2\pi \delta(\vec p-\vec p)(2\pi)^3
(-i) \frac12 \bar v^j_{p\si'} \left\{ 
M_i \frac{\Ga_j}{\pi}\frac{-i/2\cdot g_{abs}}{(\sqrt{\vec p\,{}^2
       +M_i^2}-\sqrt{\vec p\,{}^2+M_j^2})^2+\Ga_j^2} 
\right. 
\\[6mm] \di \hspace{30mm} 
+ \left. M_j
    \frac{\Ga_i}{\pi}\frac{-i/2\cdot g_{abs}}{(\sqrt{\vec p\,{}^2
       +M_j^2}-\sqrt{\vec p\,{}^2+M_i^2})^2+\Ga_i^2}
\right\} v_{p\si}^i
\end{array}
\eeq

As mentioned already, $\delta(\vec p-\vec p)(2\pi)^3$ 
represents the volume element $V$.
Similarly the time of interaction also appears as multiplicative factor.
\beq  \di
J_{ij}=\int\limits_{-\infty}^{\infty}  dt\,
e^{-i\sqrt{\vec p\,{}^2+M_i^2}t-\Gamma_i|t|}
 e^{i\sqrt{\vec p\,{}^2+M_j^2}t-\Gamma_j|t|}
=2\pi\frac{(\Ga_i+\Ga_j)/\pi}{(\Ga_i+\Ga_j)^2+
(\sqrt{\vec p\,{}^2+M_i^2}-\sqrt{\vec p\,{}^2+M_j^2})^2}.
\eeq

The new feature of this calculation is the presence of the masses
and the decay widths in the S--matrix.  
The new terms define the time intervals when the interaction takes
place.  To obtain the transition amplitudes per unit volume and unit
time we must divide by the factor $V\cdot J_{ij}$ where $V$ is the
volume element and $J_{ij}$ the time interval of the interaction.
In other words, we shall work with the ``$T$--Matrix'' defined 
through the equation 
\begin{equation}
S_{fi} = 1 + i \,\, T_{fi}(2\pi)^4 \delta^4(p_f-p_i)\, 
\end{equation}
where the energy conserving $\delta$--function will be substituted
by the expression
\begin{equation}
(2\pi)\delta(E_f-E_i) \to \frac {2(\Gamma_i+\Gamma_j)}
    {(E_f-E_i)^2 + (\Gamma_f +\Gamma_i)^2}\, .
\end{equation}

One easily arrives at
\begin{eqnarray}
T_{ij} = \frac{S_{ij}^{(I)}+S_{ij}^{(II)}}
{V\cdot J_{ij}} & = \frac{(-1)}{2}\frac{1}{16\pi}
& \left( h_{\al\,i}^*h_{\al\,j}\bar u^i_{p\si} \right.
\left[ M_i\frac{\Gamma_j}{\Gamma_i+\Gamma_j}\, P_R + M_j
         \frac{\Gamma_i}{\Gamma_i+\Gamma_j}\, 
             P_L \right] u_{p\sigma'}^j\nonumber \\
& & \left. +h_{\alpha i}h_{\alpha_j}^* \bar{v}_{p\sigma'}^j
  \left[ M_j\frac{\Gamma_i}{\Gamma_i+\Gamma_j}\, P_R + M_i
      \frac{\Gamma_j}{\Gamma_i+\Gamma_j}\, P_L \right] v_{p\sigma}^i\right).
\end{eqnarray}
Notice the S--matrix element (17), (18) vanish in the case 
$|M_i - M_j| \gg \Gamma_{ij}$ since the Lorentzian representation of
the Delta--function (21) tends to zero.  It is easy to see that the
same occurs for the transition probability
\begin{displaymath}
\frac{w(N_i\to N_j)}{V\cdot J_{ij}} =  
 |T_{ij}|^2\cdot (2\pi)^4 \cdot \delta(\vec{p}-\vec{p}\,') \cdot
   \frac{(\Gamma_i + \Gamma_j)/\pi}{(E_f-E_i)^2 + (\Gamma_f + \Gamma_i)^2}\,.
\end{displaymath}
Let us summarize our results.  We have shown that neutrinos of different
generations can mix through one--loop diagrams even if they have different
masses.  The physical reason of this possibility is related to the final
widths of neutrinos :  their mass shells are ``fuzzy'' and energy is 
conserved with the accuracy comparable to their width.  The closer the masses
of neutrinos, the larger is the transition $N_i\to N_j$ probability; and
for $|M_i - M_j| \gg \Gamma_{i(j)}$ the mixing has no physical meaning.

As we already know, neutrino mixing provides an indirect contribution
to leptogenesis.  As it was previously done in \cite{eap,weiss}, we 
are working in terms of an effective Hamiltonian, or, what is the same,
in terms of an effective mass matrix.  From the transition matrix elements
and the mass term in eq.\ (1) we can now proceed to calculate an effective
mass matrix and the physical states.

We consider the expression for $T_{ij}$ as the first order contribution
for mass matrix elements.  One should notice that the contributions are
different for $R$ and $L$ parts of our 4--spinor $N_i$ in eq. (2), i.e.\ they
are different for $N_{Ri}$ and $(N_{Ri})^c$.  Evidently, an overall
factor of $(-i)$ and the spinors originate from the definition of the 
S--matrix and do not appear on the mass matrix.  A factor of 1/2 is
also omitted in order to be consistent with the Lagrangian in eq. (1),
which has 1/2 in front of the mass term.

So the corrections to the $M_i\,\overline{(N_{Ri})^c}\,N_{Ri}$ mass term
(see Lagrangian in the form (3)) are
\begin{displaymath}
H_{ij} = \frac{(-i)}{2}\, \frac{1}{8\pi} 
    \left( h_{\alpha i}^*\, h_{\alpha j} \, 
      \frac{M_i\Gamma_j}{\Gamma_i+\Gamma_j}\, +\, 
       h_{\alpha i}\, h_{\alpha j}^* \, 
        \frac{M_j\Gamma_i}{\Gamma_i+\Gamma_j} \right)
\end{displaymath}
and the corrections to the $M_i\,\overline{N_{Ri}}\,(N_{Ri})^c$ mass term
are 
\begin{displaymath}
\tilde{H}_{ij} = \frac{(-i)}{2}\, \frac{1}{8\pi} 
    \left( h_{\alpha i}^*\, h_{\alpha j} \, 
      \frac{M_j\Gamma_i}{\Gamma_i+\Gamma_j}\, +\, 
       h_{\alpha i}\, h_{\alpha j}^* \, 
        \frac{M_i\Gamma_j}{\Gamma_i+\Gamma_j}\right)
\end{displaymath}
When $\Gamma_i=\Gamma_j$, this result coincides with those obtained in ref.
\cite{weiss}.  Our account of finite widths of Majorana neutrinos slightly
corrects the result which is not very essential.
What is important, is that the definition in eq.
(\ref{neutrino}) enables one to give 
physical meaning to the mixing of particles with different
masses.


\section{Effective Contribution to the Mass Term and Physical Neutrino
States.}
We shall work in a vector space with four basis vectors
($N_{R1}^c,\, N_{R1},\, N_{R2}^c,\, N_{R2}$) which are the states
occurring in the interaction term of eq.\ (1).   The transitions
among these states introduce an effective matrix
\begin{equation}
{\sl{m}} = \left( \begin{array}{cccc}
    0 & M_1+H_{11} & 0 & H_{12}\\
    M_1 + H_{11} & 0 & \tilde{H}_{12} & 0\\
    0 & H_{12} & 0 & M_2 + H_{22}\\
    \tilde{H}_{12} & 0 & M_2 +H_{22} & 0 \end{array}\right)
\end{equation}
with 
\begin{eqnarray}
H_{ij} & = & 2 \left[ h_{\alpha i}^* h_{\alpha j} \frac{M_i\Gamma_j}
              {\Gamma_i+\Gamma_j} + h_{\alpha i} h_{\alpha j}^* 
               \frac{M_j\Gamma_i}{\Gamma_i+\Gamma_j} \right]
                \left( -\frac{i}{2}\, g_{ab}\right)\\
\tilde{H}_{ij} & = & -H_{ij}^*\quad\quad\quad {\rm{and}}
                     \quad\quad\quad g_{ab} = \frac{1}{16\pi}
\end{eqnarray}
The $2\times 2$ matrices in the upper left and lower right corners
of eq.\ (23)
are lepton--number violating but flavor conserving, while the 
$2\times 2$ matrices along the off--diagonal are lepton-- and 
flavor--number violating.  All terms should be present and we 
found no way to reduce it to a $2\times 2$ matrix.  It remains to
diagonalize the matrix and find the wavefunctions. It is instructive
to present a perturbative solution of the problem and then discuss
the exact solution.

We split the mass matrix into the dominant term and a perturbation
\begin{displaymath}
{\sl{m}} = H_0 + \lambda V
\end{displaymath}
with
\begin{equation}
H_0 = \left( \begin{array}{cccc}
              0 & M_1+H_{11} & 0 & 0 \\
              M_1+H_{11} & 0 & 0 & 0 \\
              0 & 0 & 0 & M_2+H_{22} \\
              0 & 0 & M_2+H_{22} & 0 \end{array}\right) 
\end{equation}
and
\begin{equation}
\lambda V = \left( \begin{array}{cccc}
                    0 & 0 & 0 & \lambda H_{12} \\
                    0 & 0 & -\lambda H_{12}^* & 0 \\
                    0 & \lambda H_{12} & 0 & 0 \\
                    -\lambda H_{12}^* & 0 & 0 & 0 \end{array}\right)
\end{equation}

We introduce a small parameter $\lambda$ in order to keep track of
the perturbative corrections and at the very end set $\lambda = 1$.
Applying perturbation theory we find the eigenfunctions to 
$0(\lambda^2)$ 
\begin{equation}
U_1 = \left( \begin{array}{c}
             1 \\[1.5mm] 1 \\[1.5mm] 
	     X \\[1.5mm]
	     Y \end{array} \right)
\quad\quad\quad {\rm{and}} \quad\quad\quad
U_2 = \left( \begin{array}{c}
	     X'\\[1.5mm]
	     Y'\\[1.5mm]
	     1 \\[1.5mm] 1\end{array}\right)
\end{equation}
with the eigenvalues $\lambda_1 = M_1-i\frac{\Gamma_{11}}{2}\,\,\,$,
$\lambda_2 = M_2 -i\frac{\Gamma_{22}}{2}$, and $H_{ii} = -\frac{i}{2}
\Gamma_{ii}$.  They have the
time dependence $U_i\, e^{-i(M_i-i\frac{\Gamma_{ii}}{2})t}$, with
$i=1$ and 2 which is consistent with the time dependence introduced
in eq.\ (6).  The physical states are
\begin{eqnarray}
\psi_1 & = & \frac{1}{\sqrt{N}} \left[ |N_{R1}\rangle +|N_{R1}^c\rangle
            + X\, |N_{R2}^c\rangle + Y\, |N_{R2}\rangle \right]\\
\psi_2 & = & \frac{1}{\sqrt{2}} \left[ |N_{R2}\rangle +|N_{R2}^c\rangle
            + X'\, |N_{R1}^c\rangle + Y'\, |N_{R1}\rangle \right]
\end{eqnarray}
with $X = \frac{H_{12} M_1 + H_{12}^* M_2}{M_1^2 - M_2^2}$,
 $\,\,Y = \frac{H_{12} M_2 + H_{12}^* M_1}{M_1^2 - M_2^2}$ 
and similar formulas for $X'$ and $Y'$.

The results are the same as in ref.\ [5, 6] with an additional dependence
on the widths.  The interesting result is that the definition in eq.\ (6)
makes it possible to give a physical meaning to the mixing of particles
with different masses.

We can finally calculate the asymmetry for the decay of each state.
The width for the decay of $\psi_1$ into leptons is    	             
\begin{equation}
\Gamma(\psi_1\to \ell+\ldots ) \propto \sum_{\alpha}| h_{\alpha 1} +
        h_{\alpha_2}\,Y\,|^2
\end{equation}
and into antileptons
\begin{equation}
\Gamma(\psi_1\to \bar{\ell} +\ldots ) \propto\sum_{\alpha}
          |h_{\alpha_1}^* + h_{\alpha_2}^*\,X\,|^2\, .
\end{equation}

The lepton asymmetry, defined as 
\beq 
\de_1=\frac{\Ga_{\Psi_1\to l\phi^+}-\Ga_{\Psi_1\to 
        l^c\phi}}{\Ga_{\Psi_1\to l\phi^+}+\Ga_{\Psi_1\to l^c\phi}}, 
\eeq
and it is straightforward to calculate it 
\begin{equation}
\delta_1 = \frac{1}{8\pi}\,\,
           \frac{M_1\,M_2}{M_2^2-M_1^2}\,\,
	   \frac{Im\, (h_{\alpha 1}^*\, h_{\alpha 2})^2}{|h_{\alpha 1}|^2
            +|h_{\alpha 2}\, X|^2 +Re\, h_{\alpha 1}^*h_{\alpha 2}
               (Y+X^*)}\, .
\label{asym}
\end{equation}
In the above calculation we considered the case $|M_2-M_1|\gg H_{12}$.
In the case that the two Majorana neutrinos are nearly degenerate we
must diagonalize exactly the matrix in eq.\ (23) and then we recover
the resonance phenomenon introduced in eq.\ (17) of ref.\ \cite{weiss}.

Our study so far considered the neutrinos as free particles.  In reality,
however, they are in a backgound of fields interacting many times 
with the other particles.  These interactions at a finite temperature
can modify their masses and widths \cite{miele,fred}.  
In the present analysis we have
found that within the range of validity of eqs.\ (18) and (19) the 
$T$--matrix in eq.\ (22) is a smooth function of masses and the widths,
and for this reason we can introduce the matrix elements in the Boltzmann
equations in order to study the development of the asymmetry. This
approach was followed in ref.\ \cite{flanz}.

\section*{Summary}
We have shown that finite widths of Majorana
neutrinos play a principal role in producing  
neutrino mixing via self--energy diagrams. In our approach the widths 
of neutrinos are treated as ``fuzzy'' states on their mass shell 
and the mixing  is understood as an ``overlap'' of two fuzzy mass shell
states.
For the mathematical realization of these ideas we changed 
phenomenologically
the definition of asymptotic neutrino states, including
their widths as in eq.\ (6).  Concerning the results, we showed that
these changes  do not 
influence significantly the expressions for the effective
mass matrix and for the asymmetry generated in the case
of small mass difference. For large mass differences the asymmetry
is suppressed.

\section*{Acknowledgement}
One of us (O.L.) expresses her thanks to the members of the chair
Theoretical Physics III of the University of Dortmund for their 
hospitality, where the largest part of this work
was completed, and to Prof. G.\ Vereshkov for useful discussions.
This work was supported in part by the Bundesministerium f\"ur Bildung,
Wissenschaft, Forschung und Technologie, Bonn (05 HT9PEB 8), and in
part by NATO Collaborative Research Grant No. 97 1470.

\end{document}